\documentstyle[preprint,revtex]{aps}
\begin{document}
\draft
\begin{title}
Comment on ``Theory of Impure Superconductors:\\
Anderson versus Abrikosov and Gor'kov''
\end{title}

\author{R. J. Radtke}
\begin{instit}
Department of Physics and the James Franck Institute,\\
The University of Chicago,
Chicago, Illinois  60637
\end{instit}

\begin{abstract}
In a recent article, Kim and Overhauser have found fault with the
theory of impure superconductors by Abrikosov and
Gor'kov and have proposed an alternative formalism based on
Green's functions which are not derivable from a Dyson equation.
Although the corrections to the Abrikosov-Gor'kov theory found by
Kim and Overhauser are correct for non-retarded interactions,
I argue that these corrections do not
appear when one treats realistic retarded interactions within the
field-theoretic approach of Abrikosov and Gor'kov.
Direct numerical computation of the impurity-induced
suppression of the superconducting transition temperature $T_c$ for both
retarded and non-retarded interactions illustrates these points.
I conclude that Abrikosov-Gor'kov theory applied to physical
electron-electron interactions yields a tractable formalism which
accurately predicts the effects of magnetic
and non-magnetic impurities on $T_c$.
\end{abstract}

\narrowtext

A recent paper by Kim and Overhauser (KO) \cite{KO} discusses the effect
of non-magnetic impurities on the critical temperature $T_c$ of a model
superconductor mediated by a non-retarded BCS interaction
within the context of the theories of Abrikosov and
Gor'kov (AG) \cite{AGnor,AGmag} and Anderson \cite{Anderson}.
Literal evaluation \cite{KO} of the AG theory leads to the conclusion that
$T_c$ should be strongly suppressed by these impurities, in contradiction
to experiment.
While there are also corrections to Anderson's theory, a perturbative
calculation of these corrections leads to the expected conclusion
that the $T_c$ for normal metals
is insensitive to small concentrations of normal impurities.
In order to reconcile these two approaches, KO propose
the use of ``projected'' Green's functions, which yield a $T_c$ that is
rigorously unaffected by normal scatterers as long as the usual BCS
gap equation is valid.
``Anderson's Theorem'' is thus regained.

The use of non-retarded interactions to model the behavior of
superconducting \cite{BCS,AFSF} and superfluid \cite{He} systems
has produced a variety of important physical insights.
The temptation, therefore, is to conclude from the KO results that
the entire theory of impure superconductors based on the formalism
created by Abrikosov and Gor'kov \cite{AGnor,AGmag} is invalid.
In this comment, I will demonstrate that, while the corrections to
the Abrikosov-Gor'kov theory derived by Kim and Overhauser are
correct for a non-retarded interaction,
they do not imply that the conventional treatment of
impurities in boson-mediated superconductors needs to be modified
\cite{AGD}.

The fundamental point is that all known superconducting and superfluid
Cooper pairing is mediated by retarded interactions, whether they be
phonons in conventional superconductors, paramagnons in $^3$He, or
possibly antiferromagnetic spin fluctuations in the heavy fermion
and cuprate superconductors.
As such, the pairing potential is
frequency-dependent and is nonzero only up to some finite frequency
$\omega_D$, but is unrestricted in wavevector space.
Conventional BCS models approximate this full interaction with
a frequency-independent pairing potential cut-off in wavevector
space.
While a useful toy model for pure superconductors, this approach
leads to the unphysical results obtained by KO when applied to
a dirty superconductor.

The most tractable way to treat a fully retarded interaction is
to apply the field-theoretic approach of Abrikosov and Gor'kov as
contained in strong-coupling Eliashberg theory \cite{AM}.
This formalism is a mean-field theory in which the pairing boson is
treated in the single-exchange-graph approximation and the impurities
are included by a self-consistent perturbation theory.
The advantages of this approach are that it can readily treat
retarded interactions, it is easily generalized to include strong
electron-boson coupling, and it may be cast in the form of a Dyson
equation so that the equations can be solved numerically if not
analytically.

Most importantly, however, this formalism produces results
consistent with experiment when realistic retarded interactions
are employed.
The usual equations of Eliashberg theory require that
the calculated $T_c$ be unaffected by
the presence of non-magnetic impurities \cite{AM}.
Hence, Anderson's Theorem \cite{Anderson} is obeyed.
The slight sensitivity of $T_c$ to normal defects in real materials
can be attributed to anisotropy in the
gap function, structure in the density of states, a finite band
width, or the change of the pairing potential and the band structure
due to the presence of the impurities.
All but the last of these
have been successfully incorporated into Eliashberg theory and used
to quantitatively describe the
measured suppression of $T_c$ by impurities in the conventional
\cite{AM,Tsuneto,MK} and A15 \cite{Schachinger,Pickett} superconductors.

As a concrete example of the effect retardation can have on the
response of a superconductor to impurities, Fig. 1 shows
the calculated suppression of $T_c$ by normal and magnetic impurities
for both a non-retarded pairing potential given by the BCS model
used in KO and a
retarded pairing potential modeled by Einstein phonons.
In all cases, the
transition temperature of the pure material $T_{c0}$ was set to 7.2 K,
$\omega_D$ = 88 K, and $E_F$ = 9.5 eV in order to model Pb
\cite{AandM,note}.
One can see that the non-retarded interaction shows a strong suppression
of $T_c$ due to normal impurities, in agreement with the analysis of
Kim and Overhauser.
However, the retarded interaction produces no such decrease in $T_c$;
in fact, to the numerical precision to which the Eliashberg equations
were solved (about 2 \%), $T_c$ is completely unchanged by
the presence of normal impurities \cite{Marsiglio}.
In contrast, one can see from the inset to Fig. 1 that the
Abrikosov-Gor'kov theory of magnetic impurities \cite{AGmag} is
qualitatively but not quantitatively
obeyed for {\it both} the retarded and non-retarded
pairing potentials.
Essentially all of these results have appeared in the literature
in one form or another; the numerical calculations in Fig. 1
simply present these
results in a form that allows ready comparision between retarded and
non-retarded interactions and magnetic and non-magnetic impurities.

If one still insists on using a non-retarded
interaction and would like to keep Anderson's Theorem, there are three
possible approaches, the first two of which were pointed out by KO.
The first is to transform the full electron-electron interaction
from wavevector space into the space of exact eigenstates of the
non-interacting, impure electron gas, apply the cut-off to the
exact eigenenergies,
and then use this potential in the appropriately modified BCS
equations.
The second method is to
use these exact wave functions to construct electronic Green's
functions which
contain only those wavefunctions whose eigenenergies are within
$\omega_D$ of the Fermi energy and then to insert them into the BCS gap
equation.
The final option is to transform the Matsubara sum in the gap
equation into a frequency integral and to truncate the frequency
integral at $\omega_D$ \cite{MK}.
Neither of the first two choices allow a computation of
the properties of real materials, whereas the last is
easily evaluated and is directly
related to the actual retarded nature of the interaction which is
neglected in this approximation.
Clearly, the last option is preferred.

In conclusion, although the unphysical results obtained from
the application of the Abrikosov-Gor'kov theory to
non-retarded interactions pointed out
by Kim and Overhauser are present, the AG theory yields the correct
behavior when used with a physical retarded interaction.
If one still wishes to consider the toy model of a non-retarded
interaction in an
impure superconductor, the most useful procedure
is to apply the BCS cut-off in frequency rather than in wavevector
space \cite{MK}.

\acknowledgements

This work was supported by NSF-STC-9120000 and NSF-DMR-MRL-8819860.
I would also like to acknowledge helpful conversations with
K. Levin and to thank A. J. Leggett for calling attention to this
problem.

\figure{Critical temperature relative to the critical temperature
in the absence of impurities $T_c / T_{c0}$ vs. impurity scattering
rate $\tau_{imp}^{-1}$ in meV for a non-retarded, BCS superconductor with
normal (long dashed line) and magnetic (dot dashed line) impurities
and a retarded, Einstein-phonon-mediated superconductor with
normal (short dashed line) and magnetic (triple-dot dashed line)
impurities.
In all cases, the coupling strength is chosen to give $T_{c0}$ = 7.2 K,
the BCS cut-off and the Einstein phonon frequency are 88 K, and the
Fermi energy is 9.5 eV in order to model Pb.
Inset:  $T_c / T_{c0}$ vs. the pair-breaking parameter
$\alpha = 1 / (1 + \lambda) \tau_{imp} T_{c0}$ for the magnetic
impurity data from the main figure.  The solid line is the
predicted curve from the Abrikosov-Gor'kov theory [consult Ref.
\cite{AGmag} and \cite{AM} for details]. }


\begin{references}
\bibitem{KO}Yong-Jihn Kim and A. W. Overhauser, {\it Phys. Rev. B}
{\bf 47}, 8025 (1993).

\bibitem{AGnor}A. A. Abrikosov and L. P. Gor'kov,
{\it Zh. Eksp. Teor. Fiz.} {\bf 35}, 1558 (1958) and
{\it ibid.} {\bf 36}, 319 (1959) [{\it Sov. Phys.--JETP} {\bf 8},
1090 (1959) and {\it ibid.} {\bf 9}, 220 (1959)].

\bibitem{AGmag}A. A. Abrikosov and L. P. Gor'kov,
{\it Zh. Eksp. Teor. Fiz.} {\bf 39}, 1781 (1960)
[{\it Sov. Phys.--JETP} {\bf 12}, 1243 (1961)].

\bibitem{Anderson}P. W. Anderson, {\it J. Phys. Chem. Solids} {\bf 11},
26 (1959).

\bibitem{BCS}J. Bardeen {\it et al.}, {\it Phys. Rev.} {\bf 108},
1175 (1957).

\bibitem{AFSF}K. Miyake {\it et al.}, {\it Phys. Rev. B} {\bf 34},
6554 (1986); D. J. Scalapino {\it et al.}, {\it Phys. Rev. B} {\bf 34},
8190 (1986) and {\it ibid.} {\bf 35}, 6694 (1987).

\bibitem{He}P. W. Anderson and W. F. Brinkman, {\it Phys. Rev. Lett}
{\bf 30}, 1108 (1973) and Sadao Nakajima, {\it Prog. Theor. Phys.}
{\bf 50}, 1101 (1973).

\bibitem{AGD}This conclusion appeared in print some time ago in
A. A. Abrikosov {\it et al.}, {\it Methods of
Quantum Field Theory in Statistical Physics} (Dover Pulications, Inc.,
New York, 1963), p. 337.

\bibitem{AM}See P. B. Allen and B. Mitrovi\'{c} in {\it Solid State Physics},
Vol. 37, edited by H. Ehrenreich, F. Seitz, and D. Turnbull
(Academic, New York, 1982) and references therein.

\bibitem{Tsuneto}Toshihiko Tsuneto, {\it Prog. Theor. Phys.} {\bf 28},
857 (1962).

\bibitem{MK}D. Markowitz and L. P. Kadanoff, {\it Phys. Rev.} {\bf 131},
563 (1963).

\bibitem{Schachinger}E. Schachinger {\it et al.},
{\it J. Phys. F} {\bf 12}, 1771 (1982).

\bibitem{Pickett}W. E. Pickett, {\it Phys. Rev. B} {\bf 26}, 1186 (1982).

\bibitem{AandM}These paremeters were taken from the tables in
N. W. Ashcroft and N. D. Mermin, {\it Solid State Physics}
(W. B. Saunders Company, Philadelphia, 1976).

\bibitem{note}For the Einstein-phonon-mediated superconductor,
$\omega_D$ is taken to be the phonon frequency, while for the BCS model,
it is the cut-off in wavevector space.
Also, the density of states is taken to be flat
with a band width $2 E_F$ as in Ref. \cite{KO}
and is fully accounted for when computing $T_c$.
The non-retarded equations are those appearing
in Ref. \cite{Marsiglio} generalized to include a cut-off less than
the band width and a fully self-consistent treatment of the pairing
interaction; the retarded equations are those from Refs.
\cite{Schachinger} and \cite{Pickett} modified for a flat density
of states.
Finally, he chemical potential is chosen so that the interacting band is
half-filled.

\bibitem{Marsiglio}This result is not as trivial as it sounds.
F. Marsiglio, {\it Phys. Rev. B} {\bf 45}, 956
(1992) has shown that $T_c$ is sensitive to normal impurities
if the band width is on the order of or smaller than $T_c$.
The band width here is too large to show such effects, but they do
appear at smaller band widths, although the suppression of $T_c$ is
not as strong as in the non-retarded calculations.

\end{references}
\end{document}